\documentstyle[12pt,mont98_desy,amssymb,epsfig]{article}

\def\be{\begin{equation}}
\def\ee{\end{equation}}
\def\bea{\begin{eqnarray}}
\def\eea{\end{eqnarray}}

 \def\pb{\vec{p}_{op}} 
 \def\pd{\vec{\pi}_{op}}\def\pk{\vec{\pi}}

 \def\Haq{J_{op}}  \def\ha{J}

 \def\Phb{\phi_{op}}

 \def\Ab{\vec{A}_{op}}

 \def\Bb{\vec{B}_{op}}

 \def\Eb{\vec{E}_{op}}

 \def\Psb{\vec{\sigma}_{op}}

\begin{document}

\title{I.~~ELECTRON AND PROTON SPIN POLARISATION IN STORAGE RINGS 
          --- AN INTRODUCTION 
\footnote{Updated version of a contribution to the proceedings of the
 15th ICFA Advanced Beam 
Dynamics Workshop: ``Quantum Aspects of Beam Physics'', Monterey, 
California, U.S.A., January 1998.
Also in DESY Report 98--096, September 1998.}
}
\author{D. P. BARBER}

\address{Deutsches Elektronen--Synchrotron, DESY, \\
 22603 Hamburg, Germany. \\E-mail: mpybar@mail.desy.de}

\maketitle
\abstracts{ This article provides a unified introduction to the 
theory of electron and proton spin polarisation in storage rings
and it provides  a common starting point for the
written versions of the four talks that I gave at Monterey. }

\section{Foreword}
Each of the four talks that I gave at Monterey had to do with spin 
polarisation 
in storage rings and accelerators and in each talk I covered the relevant 
and necessary aspects of the theory. Indeed, three of the talks essentially
dealt only with theory and there was considerable repetition of the basics.  
If the written versions were to reflect the talks as I delivered them, there
would again be repetition but there would also be an apparent lack of 
connection between those topics which were specific to each
talk. Thus a reader who survived reading all four articles might 
still not have a solid view of the connections between the concepts
covered. So it seems appropriate to provide a common introduction to 
the theory. That is the burden of this article. This also provides a suitable
opportunity to present a synthesis of the various ways of describing the
competition between polarisation build--up and depolarisation for 
electrons that I have come  across or contributed to over the last decade.
Moreover it is an opportunity to lay to rest  some confusions that 
have crept into the subject. 
Owing to space limitations I will not attempt to maintain a high degree 
of mathematical rigour but aim instead to impart a feeling for the 
issues and for our current level of understanding.
I shall refer to this article as Article I. 


The written versions of the talks themselves
will be referred to as Articles II, III, IV, and V 
as follows:
\begin{itemize}
\item[II]
Longitudinal electron spin polarisation at $27.5~GeV$ in HERA. \\
 ( D.P.~Barber  {\it for the HERA Polarisation Group } )
\item[III]
The permissible equilibrium polarisation distribution in a stored proton beam.
( D.P.~Barber,  K.~Heinemann, M.~Vogt and  G.H.~Hoffst\"atter )
\item[IV]
Unruh effect, spin polarisation and the Derbenev-Kondratenko \\
formalism.
( D.P. ~Barber )
\item[V]
The semiclassical FW transformation and the derivation 
of the Bloch equation for spin--1/2 polarised beams 
using Wigner functions. \\
( K.~Heinemann~ and~  D.P.~Barber )
\end{itemize}

\section{Introduction}
Spin behaviour in the electromagnetic guide fields of storage rings is
dominated by two effects:
\begin{itemize} 
\item Spin precession 
\item Spin flip due to synchrotron radiation emission 
\footnote{But it will become clear later that the distinction 
between the two can
become blurred in storage rings. Indeed resonant spin flip in nuclear 
magnetic resonance experiments can be viewed either as flip due to photon
absorption or precession by $\pi$ around an effective horizontal field.}.
\end{itemize}
In existing proton rings  and those that will be built in the foreseeable
future only spin precession is of significance since the synchrotron 
radiation power emitted by protons is negligible. 
However, as pointed out by Sokolov and Ternov in 1964, radiative 
spin flip can, 
for electrons, lead to a build up of polarisation \cite{st64}. This
phenomenon is then commonly known as the Sokolov--Ternov (ST) 
effect \footnote{In these articles statements made about electrons will 
also apply to positrons except for appropriate trivial sign changes 
in mathematical expressions.}.
At the time of writing, the only known practical way of obtaining a 
stored polarised proton beam is to inject a prepolarised beam provided by
a suitable source \cite{zel97} and then accelerate it.
Nevertheless another method has been suggested and I will comment on that
in Article III.

In the remainder of this article I will provide a unified overview 
of spin precession and spin flip and show how to arrive at an efficient 
description of their combined effect.

\section{Spin precession}
Spin precession for particles travelling in the electromagnetic fields 
in storage rings is most conveniently described in terms of
the Thomas--Bargmann--Michel--Telegdi (T--BMT) equation 
\cite{th27,bmt59,jackbook}:

\begin{eqnarray}
     \frac{d}{dt}
               \ \vec{S}
                          &=&
          \vec{\Omega}\wedge
                                \vec{S}
\end{eqnarray}
where $\vec S$ is the 3--vector describing spin in the centre of mass frame
and
\begin{eqnarray}
   \vec{\Omega}
 =       \frac{e}{m c}
            \left[
        -\left(\frac{1}{\gamma}+a\right) \vec{B}
        +\frac{a\gamma}{1+\gamma} \frac{1}{c^{2}}
     (\dot{\vec{r}} \cdot \vec{B})
      \dot{\vec{r}}
         + \frac{1}{c} \left(a+\frac{1}{1+\gamma}\right)
     (\dot{\vec{r}}\wedge{\vec E})
                  \right] \, .
\end{eqnarray}
The vector $\vec{B}$ is the magnetic field, 
$\vec{E}$ is the electric field and $\gamma$ is the Lorentz factor.
The vectors
$\vec{r}$ and $\dot{\vec r}$ are the position and velocity and evolve 
according to the Lorentz equation. The quantity
$a=(g-2)/2$ is the gyromagnetic anomaly.
For electrons $a \approx 0.0011596$ and for protons $a \approx 1.7928$. 
The other symbols  used here and elsewhere have their usual meanings. 
The derivations of the T--BMT equation by its authors were purely classical
in spirit. The derivation by BMT was based on the requirements of 
relativistic covariance. However, Thomas combined conventional 
notions of spin precession with the relativistic effect now called 
Thomas precession \cite{th27,jackbook}
\footnote{Thomas also provided covariant forms for his equation.}.
 Note that Eqs.~(1) and (2) 
reduce smoothly to the usual nonrelativistic limit.
To obtain a clearer view of the implications of the T--BMT equation one can 
rewrite it in terms of the field components perpendicular and parallel to the
orbit:
\begin{eqnarray}
&& \frac{d \vec {S}}{dt}   =  \frac{e \vec {S}}{m c \gamma} \wedge
(( 1 + a) \vec {B}_{\|} + (1 + a \gamma) \vec {B}_{\perp}) \nonumber  \\
&& \qquad = 
 \frac{e \vec {S}}{m c} \wedge
(( \frac{g}{2 \gamma}) \vec {B}_{\|} + 
(\frac{1}{\gamma} -1 + \frac{g}{2}) \vec {B}_{\perp}) \,  ,
\end{eqnarray}
where for this part of the discussion the effect of electric fields
has been ignored. Eq.~(3) shows that for motion perpendicular to the
 field, 
the spin precesses around the field at a rate $1 + a\gamma$ faster than the
corresponding rate of orbit deflection:
\begin{equation}
\delta\theta_{spin} = (1 + a \gamma) \delta\theta_{orbit}
= a \gamma \delta\theta_{orbit} + \delta\theta_{orbit}
\end{equation}
in an obvious symbolic notation.
 This precession rate is strongly 
influenced by the Thomas precession. This is contained in the term
$1/\gamma -1$. For electrons ($g \approx 2$) the total precession is
strongly suppressed. For protons ($g \approx 5.58$) the 
relative suppression is much weaker.

However,  `spin' is a purely quantum mechanical concept. Moreover, we are not
working in a regime where electron--positron  creation and annihilation 
are important. Thus a two--component description of spin should 
suffice and  one should therefore look for a Foldy--Wouthuysen 
transformation (Article V) of the
Dirac Hamiltonian (containing a `Pauli' term for the anomalous magnetic moment)
appropriate for the semiclassical regime of a storage ring.
By `semiclassical' I mean that for the high energies involved it should only
be necessary to keep terms up to first order in $\hbar$.
A Hamiltonian of the required type 
was already written down in 1973 by Derbenev and Kondratenko (DK)
\cite{dk73} and takes the form
\footnote{The subscript `op' is to remind the reader that we are dealing with
operators. In this case they operate on two--component wavefunctions.
The fields in $h_{op}^{dk}$ are external fields. 
The derivation of this Hamiltonian is the subject of Article V.}
%
%
%
%
%
%
%
%
\begin{eqnarray}
&& h_{op}^{dk}  =  h_{op,orb}^{dk}  
+ \frac{\hbar}{2}\Psb\cdot
\vec{\Omega}_{op} \; ,
\end{eqnarray}
where: 
\begin{eqnarray}
&& h_{op,orb}^{dk}  = \Haq + e \Phb  \; ,
\end{eqnarray}
and:
\begin{eqnarray}
&& \vec{\Omega}_{op}  {=}
-\frac{e}{2 m c} (\frac{m c^2}{\Haq}+
\frac{g-2}{2}) \Bb
+\frac{e c (g-2)}{4  m}
\frac{1}{\Haq (\Haq+m  c^2)} 
\pd (\Bb \cdot \pd)
\nonumber\\&&
+\frac{e}{2  m}
\biggl(\frac{g-2}{2 \Haq} 
+ \frac{m  c^2}{\Haq (\Haq+m c^2)}\biggr)
(\pd\wedge\Eb)
 + {\rm h. c.} \;  , 
\end{eqnarray}
and where $\pd$ and $\Haq$ are defined as:
\begin{eqnarray}
&& \pd = \pb  - \frac{e}{c} \Ab  \; , \qquad
\Haq = \sqrt{c^2 \pd\cdot\pd+m^2  c^4} \; .
\end{eqnarray}

Thus the DK Hamiltonian consists of a purely orbital part of zeroth order
in $\hbar$ and a spin part of first order in $\hbar$. The 
orbital part resembles the familiar form of the classical relativistic
Hamiltonian from the textbooks \cite{goldstein} and the spin part
is reminiscent of a Stern--Gerlach (SG) dipole energy term.
As will be noted in  Article V, at second order in $\hbar$ this 
Hamiltonian gains just extra  orbital terms. All in all, the DK Hamiltonian 
has a satisfying and physically transparent form. It is then no surprise that
in first order in $\hbar$
the Heisenberg equation of motion (EOM) for the kinetic momentum $\pd$ 
is the Lorentz equation with an additional term for the SG force.
It is also clear  that in first order in $\hbar$ the Heisenberg EOM  for the
spin $(\hbar/2)\Psb$ is a precession equation  with 
the same form as the T--BMT equation, Eqs.~(1) and (2), since 
the operator $ {\vec{\Omega}}_{op}$ has a structure equivalent to that of
 ${\vec{\Omega}}$ in Eq.~(2).
In a wave packet approximation and at first order in $\hbar$
the Heisenberg EOM lead to the T--BMT
equation for the expectation value 
$\langle \Psb \rangle$ ($=$ the polarisation) and 
the EOM for the expectation value $\langle{\pd}\rangle$ 
of the kinetic momentum 
of a wave packet is again the Lorentz equation  modified
by a SG term \cite{fg61}. Thus we have now put the T--BMT equation
 on a firm quantum 
mechanical footing and have shown that it is the natural outcome of a 
semiclassical approximation. Moreover (see Article V), we know how to
calculate beyond first order in $\hbar$ if necessary. Note that the
magnetic  SG 
terms differ from the familiar textbook forms for slowly moving particles
but reduce to them at low energy:
our terms contain Thomas
precession contributions so that, for example, $g/2$ is
 replaced by
$g/2 - 1 + 1/\gamma = a + 1/\gamma$. 
A detailed discussion on the SG terms
in the DK Hamiltonian and on the SG forces allowed by covariance can be found
in \cite{kh96} where the EOM are given a 
classical interpretation. See also Article III. 

The full Hamiltonian given by Derbenev and Kondratenko to include  radiation 
effects is 
\begin{eqnarray}
h_{tot}^{dk} = h_{op}^{dk} + h_{rad}^{dk} + h_{int}^{dk}
\end{eqnarray}
where $h_{rad}^{dk}$ is the Hamiltonian of the free radiation field and where
\begin{eqnarray}
h_{int}^{dk}=  e({\phi}_{rad} - \frac{\vec v}{c}\cdot{\vec A}_{rad}) + 
\frac{\hbar}{2} (\Psb\cdot
\vec{\Omega}_{rad}) \;
\end{eqnarray}
describes the particle--radiation interaction. The operator 
$\vec{\Omega}_{rad}$ has the same structure as $\vec{\Omega}_{op}$
except that the external field operators (denoted by the subscript {\it `op'})
are replaced with radiation field operators (denoted by the subscript
{\it `rad'}). 

\section{Spin distributions}
In the last section it became clear  that to first order in $\hbar$
the centres of wave packets move 
(classically) according to the usual Lorentz force modified by a SG term
 and that the accompanying $\langle(\hbar/2)\Psb \rangle$
 obeys the T--BMT equation. Thus  for many purposes 
the particles and their spins can be treated as if they are classical objects
and  we are then in a position to move beyond single particles and to
discuss classical spin and phase space distributions. Article V shows 
how to arrive at spin and particle distributions directly from 
the density operator.

To construct a classical treatment one uses the correspondences:
\begin{eqnarray}
\langle{\vec r}_{op}\rangle \rightarrow \vec r \; , \qquad
\langle{\pd}\rangle \rightarrow \pk \; , \qquad
\langle\frac{\hbar}{2}{\Psb}\rangle \rightarrow \vec {\xi} \; 
\end{eqnarray}
where $\vec \xi$  is a classical spin of length $\hbar/2$.
Then with the Hamiltonian:
\begin{eqnarray}
&& {\frak h}^{dk}  = {\frak h}_{orb}^{dk}  
+ {\vec \xi}\cdot
{\vec{\Omega}} \;  
\end{eqnarray}
with
\begin{eqnarray}
&& {\frak h}_{orb}^{dk}  = \ha + e\cdot\phi  \; 
\end{eqnarray}
and the Poisson bracket relations
\footnote{If we were working to second or higher order in $\hbar$
we would use the Moyal algebra \cite{moyal,khmoyal}. In the present case of
first order in $\hbar$ this simplifies to the Poisson  algebra.}:
\begin{eqnarray}
&& \lbrace r_{{}_{j}},p_{{}_{k}}\rbrace = \delta_{jk} \; , \qquad
  \lbrace r_{{}_{j}},r_{{}_{k}}\rbrace =
 \lbrace p_{{}_{j}},p_{{}_{k}}\rbrace =
   \lbrace r_{{}_{j}} , \xi_k\rbrace =
    \lbrace p_{{}_{j}} , \xi_k\rbrace = 0 \; ,\qquad
\nonumber\\
&& \lbrace \xi_j , \xi_k\rbrace =
    \sum_{m=1}^3 \varepsilon_{jkm}~ \xi_m
                                            \;  ,
\qquad (j,k=1,2,3) \, ,
\end{eqnarray}
and where semiclassically the ${\dot{\vec r}}$ in Eq.~(2)
equals $c^2 \pd/J$, 
the Lorentz (modified by a SG term)
 and T--BMT equations emerge from the canonical equations of motion:
\begin{eqnarray}
\dot{\vec r} = \lbrace \vec r, {\frak h}^{dk} \rbrace \; ,\qquad
\dot{\pk} = \lbrace \pk, {\frak h}^{dk} \rbrace  
                + \frac{\partial{\pk}}{\partial{t}} \; ,\qquad
\dot{\vec{\xi}} = \lbrace \vec{\xi}, {\frak h}^{dk} \rbrace \; .
\end{eqnarray}
Since storage rings and accelerators have accelerating cavities which
subject the particles to time dependent fields and since the magnet geometry 
is fixed, particle dynamics is best described in terms of the canonical 
coordinates  $\vec u = (x, p_x, z, p_z, \Delta t, \Delta E)$ 
where $x,~ p_x,~ z,~ p_z$ describe transverse motion with respect to the
curved  periodic  orbit and $\Delta t, \Delta E$ are the time delay relative
to a synchronous  particle (at the centre of the bunch) 
and the energy deviation from the energy of 
a synchronous particle respectively. The independent variable is now
the distance around the ring, $s$. There is a corresponding (classical)
Hamiltonian, correct up to first order in $\hbar$, 
\begin{eqnarray}
\tilde{h}= {\tilde{h}}_{orb} + {\vec \xi}\cdot\vec{\tilde{\Omega}} \; ,
\end{eqnarray}
which enables the EOM to be written in canonical form and this
is derived from ${\frak h}^{dk}$ 
by standard means \cite{bhr1}.
If the ring is distorted (see below),
 $\vec u$ describes the position with respect to the resulting closed orbit.

We now make the idealisation  that the beam phase space
can be described in terms of a smooth continuous density, $w(\vec u;s)$,
which is a scalar function of $\vec u$ and the azimuth $s$
\footnote{Note that in Article V the phase space density is denoted by 
`$\rho$'.}.
It is normalised to unity.
In the absence of dissipation and noise  (e.g. due to synchrotron radiation) 
and ignoring the effect of the tiny SG 
forces on the orbital motion, $w$ is constant along a phase space
trajectory and obeys a relation of the Liouville type:
\begin{eqnarray}
 \frac{\partial w}{\partial s}~ =~ \lbrace{\tilde{h}}_{orb} , 
w \rbrace
                                                                          \, .
\end{eqnarray}
If the beam is stable, i.e. if  $w$ is the same from turn to turn,
then it is periodic in $s$  
and we write it as $w_{eq}$ so that 
$w_{eq}(\vec u;s)= w_{eq}(\vec u;s+C)$, where $C$ is the
ring circumference.

Having assigned a phase space density to each point in phase space we
now assign  a polarisation  $\vec P(\vec u ;s)$ to each
point 
\footnote{ This is equivalent to associating a spin density matrix with
each point in phase space.}.
$\vec P$ is the average over particles of the  unit spins
 $2{\vec \xi}/{\hbar}$ at
 $(\vec u; s)$.
Since the T-BMT equation is linear in the spin and since in this
picture the spins at $(\vec u; s)$
all see the same $\vec{\tilde\Omega} (\vec u; s)$,~
$\vec P(\vec u; s)$ obeys the T-BMT equation
\begin{equation}
\frac{d \vec {P}}{ds}= \vec {\tilde\Omega}({\vec u}(s);s)   \wedge \vec {P}~.
\end{equation}
Because Eq.~(18) describes precession,
$|\vec P(\vec u; s)|$ is
constant along a phase space trajectory.
To make closer contact with the synchrobetatron motion, we can rewrite
Eq.~(18) as \cite{erice95,kh97}:
\begin{equation}
  \frac{\partial \vec{P}}{\partial s}~ =~
      \lbrace {\tilde h}_{orb},  \vec{P}  \rbrace
        +   \vec{\tilde\Omega}({\vec u};s)  \wedge   \vec{P}
\end{equation}
which is analogous to Eq.~(17) and assumes that $\vec P(\vec u; s)$ is 
differentiable in all directions in phase space.
Note that the polarisation of the whole
beam as measured by a polarimeter at azimuth $s$ is the average across 
phase space: 
\begin{equation}
\vec{P}_{av}(s) =  
       {\int} d^6 u~w({\vec u}; s){\vec P}({\vec u}; s) \, .
\end{equation}
If the spin distribution is stable, i.e. if $\vec P(\vec u; s)$ is the same
from turn to turn, then 
$\vec P(\vec u; s)$ not only obeys the T-BMT equation, but it is also
periodic in $s$ and we write it as ${\vec P}_{eq}$ so that
${\vec P}_{eq}(\vec u; s) = {\vec P}_{eq}(\vec u; s+C)$.
We denote the unit vector along ${\vec P}_{eq}(\vec u; s)$
by $\hat n(\vec u ;s)$
\footnote{              With respect to $\hat n(\vec u ;s)$ the spin
density matrix at ~($\vec u; s$) is diagonal.}.
This also obeys Eq.~(18) and is periodic in $s$: 
${\hat n}(\vec u; s) = {\hat n}(\vec u; s+C)$.
On the (periodic) closed orbit  
$\hat n(\vec u; s)$ becomes $\hat n(\vec 0; s)$ and we denote it by
 $\hat n_0(s)$
\footnote{{\label{foot:gdndg}}Many authors make no clear distinction between 
$\hat n$ and $\hat n_0$ and many use the symbol $\hat n$ for $\hat n_0$.
This can sometimes lead to confusion.
In particular the original symbol for 
$\partial{\hat{n}}/{\partial{\delta}}$ (section 5.3) was 
$\gamma \partial{\hat{n}}/{\partial{\gamma}}$ \cite{dk73} and some have 
erroneously understood  $\gamma \partial{\hat{n}}/{\partial{\gamma}}$  
 to mean 
$E_0 \partial{\hat{n}_0}/{\partial{E_0}}$ where $E_0$ is the design energy.}.
Obviously $\hat n_0(s)$ obeys the periodicity condition
$\hat n_0(s) = \hat n_0(s+C)$, i.e. $\hat n_0(s)$ is the ring periodic 
solution of the T--BMT equation on the closed orbit. In general it is unique. 

In the foregoing I introduced the 
{\it invariant (vector) spin field} ${\hat n}(\vec u; s)$ 
by appealing to physical intuition. The underlying assumption was that
the field ${\hat n}(\vec u; s)$, which is supposed to obey Eq.~(18) over 
the whole of the beam phase space, not only {\it exists} but is smooth
(to correspond with our expectations of the spin distribution in a real beam)
and is unique. However, the situation is not quite so simple as I will 
now explain by describing some qualitative aspects of spin motion. 

If a circular accelerator only had vertical (dipole) fields, vertical
spins would not be affected and $\hat n_0(s)$ would be vertical. Moreover,
according to Eq.~(3) a non-vertical spin would precess around $\hat n_0(s)$
$a \gamma$ times per turn with respect to the (periodic) design orbit.
I call the quantity $a \gamma$ the `naive spin tune'. It represents the natural
spin precession frequency of this simple system. It increases by one unit for
every $\approx 440~MeV$ ($\approx 523~MeV$) increase in the energy of
 electrons (protons). 
But some rings contain vertical bend magnets so that the design orbit is not
flat. The ring might also contain solenoidal fields  of particle detectors.
In these cases a periodic T-BMT solution, $\hat n_0(s)$,  
on the design orbit still exists but is no longer everywhere vertical and it is
given by the real eigenvector (with
unit eigenvalue) of the one turn (orthogonal) 3 x 3   spin transfer
matrix for this design orbit
\footnote{{\label{foot:mont}}However, for $\vec u \neq {\vec 0}$, the 
constraint 
${\hat n}(\vec u; s) = {\hat n}(\vec u; s+C)$ 
is obviously {\it not} equivalent to an analogous eigenproblem for 
${\hat n}(\vec u; s)$ since in general a spin at $(\vec u; s)$ set parallel to 
${\hat n}(\vec u; s)$ does not map into itself over one turn. Thus the 
{\it naive}  
algorithm based on a one turn map (e.g. see page 27 in \cite{mont84}) is 
incorrect; in general a `${\hat n}$' constructed in that way would not
obey the T--BMT equation everywhere along an orbit.
As a result, a `${\hat n}$' constructed in that way should not be used to 
obtain  the vector $\partial{\hat{n}}/{\partial{\delta}}$ needed, as in 
section 5.3, to describe radiative depolarisation of electrons 
(e.g. see page 52 in \cite{sylee98}). However, ${\hat n}$ can be obtained
as an eigensolution of a {\it modified} eigenproblem \cite{yok92,yok99}.
See also footnote {\it {\ref{foot:gdndg}}}.
}.
Indeed, for the HERA  electron ring (Article II) $\hat {n}_0$
is made {\it longitudinal} at the east IP by means of spin rotators.
The number of spin 
precessions around  $\hat n_0(s)$ per turn  is extracted from the complex  
eigenvalues of the matrix \cite{chao81,chao82}.
 We call this the `real spin tune'
or just the `spin tune' and denote it by $\nu_{spin}$.
In general it deviates from  $a \gamma$ 
\footnote{Actually, the complex eigenvalues only deliver the fractional
part of the spin tune. The integer part must be found by following the spin 
motion for one turn.}.

If the spin tune were an integer, the one turn matrix would be a unit matrix
and  $\hat n_0(s)$ would not be unique. This lack of uniqueness also 
manifests itself in extreme sensitivity to field errors. 
The quadrupoles and other magnets in storage rings normally have 
unavoidable small misalignements so that the periodic (closed) orbit
deviates from the design orbit. Likewise the real $\hat n_0(s)$ deviates from  
the design $\hat n_0(s)$ since a spin on the closed orbit now `sees'  
horizontal dipole
components in the quadrupoles. There is also a small shift in the real 
spin tune. The angle between the two $\hat n_0(s)$'s
is roughly proportional to the amount of closed orbit distortion. But it 
becomes very large if the design spin tune is close to an integer 
\cite{br98,bmrr85} since the spin motion is then coherent with the imperfection
fields. The spins are then said to be near an integer resonance
(sometimes called an `imperfection resonance').   

Particle bunches in storage rings have  nonzero transverse dimensions and 
energy spread and the motion of a spin, compared to that of a spin on the 
closed orbit, depends on the position in phase space via the  $\vec u$
in $\vec{\tilde\Omega} (\vec u; s)$. 
For particles circulating for many turns the total disturbance to a spin can
grow to become  very big if there is coherence between the natural 
spin motion and the oscillatory motion in the beam characterised by
the spin-orbit resonance condition:
\begin{equation}
{\nu}_{spin} = m + m_x ~Q_x + m_z~ Q_z + m_s~ Q_s
\end{equation}
where the $m$'s are integers and the $Q$'s are respectively the horizontal, 
vertical and longitudinal tunes of the orbital oscillations.

The integer resonances ($|m_{x}|+ |m_{z}|+ |m_{s}| =0$ in Eq.~(21)) can 
normally be identified with the imperfection resonances already mentioned and 
driven by the periodic imperfection fields along the closed orbit. 
We have absorbed their influence into a large deviation of $\hat n_0(s)$
from the design direction.
The spin-synchrobetatron resonances
($|m_{x}|+ |m_{z}|+ |m_{s}| \neq 0$ in Eq.~(21))
(sometimes called `intrinsic resonances') are driven by the
quasiperiodic fields seen by particles executing quasiperiodic synchrobetatron
oscillations about the closed orbit. 
The sum $|m_{x}|+ |m_{z}|+ |m_{s}|$ is called
the order of the resonance. An imperfection resonance is then a zeroth order
resonance. 
Although I have just been discussing the behaviour near resonance of 
arbitrary  
spins it should now be clear that $\hat n$, which is a special solution of 
Eq.~(18) constrained to be periodic, should, just like $\hat n_0(s)$, 
also show extreme behaviour 
near resonances. This is confirmed by the analytical structure and numerical
output from 
the algorithms used for its construction \cite{br98}.
Near integer resonances in  a distorted ring 
$\hat {n}_0$ deviates  
strongly from the nominal direction for the  perfectly aligned ring and near  
intrinsic resonances the difference 
$\hat n(\vec u ;s) - \hat {n}_0$ becomes large and increases with the
synchrobetatron amplitude $\vec u$ and with $a \gamma$
\footnote{
Note that the terminology `intrinsic' and `imperfection' 
must be used with care since synchrobetatron motion can also give rise to
zeroth order resonance phenomena \cite{mane90,mane92}.}.

For  $27.5~GeV$ electrons in HERA (see Article II) the r.m.s. angle between
 $\hat n(\vec u ;s)$ and 
$\hat {n}_0$ (obtained by averaging across phase space) 
is just a few milliradians away from  intrinsic resonances and about
 100 milliradians very near such resonances. For protons at about $800~GeV$ 
in HERA (see Article III)
on the `1--$\sigma$' torus this angle is  typically
 60 degrees unless  
Siberian Snakes are employed.
\begin{figure}[bthp]
\begin{center}
\epsfig{figure=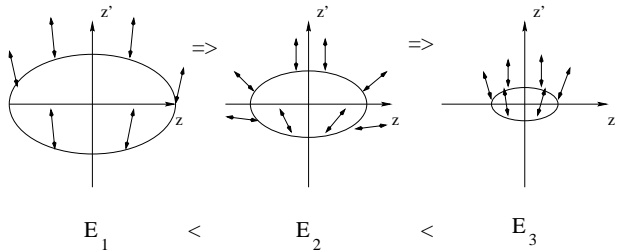,width=5cm,angle=-90}
\end{center}
\caption{\footnotesize
A typical $\hat{n}$--field at three energies, the second of which is
close to resonance.}
\label{fg:Adiab}
\end{figure}
Figure 1 depicts invariant spin fields $\hat n$ `attached' to vertical betatron
phase space ellipses for three different fixed energies but
for the same invariant vertical emittance. Other examples are given in
Article III. 

Although I introduced $\hat n$ via spin distributions, the history 
of $\hat n$  took a different  course which provides more insight into its
meaning and properties.
It was first introduced 
by Derbenev and Kondratenko \cite{dk72,dk73} in the process of obtaining 
action--angle variables for combined spin--orbit motion by `diagonalising'
the Hamiltonian in Eq.~(12) and this aspect was further illuminated in  
\cite{yok86}.
A similar approach can be used on the 
Hamiltonian in Eq.~(16) \cite{bhr1,bhrnotes,yok86,bal98}. I now give
a rough outline of the basic ideas.

It is assumed that the orbital motion is integrable and one  makes 
an $s$ dependent canonical transformation so that
$\tilde{h}_{orb}$ is replaced by 
${\bar{h}}_{orb} = \Sigma_i 2\pi {{\bar Q}_i}{\bar I}_i$ 
where the ${\bar Q}_i$ are the three orbital tunes and the 
three ${\bar I}_i$ are the
components of the orbital action vector $\vec{\bar I}$.
Then one describes the spin motion with respect to a set of orthonormal 
axes (a `dreibein')
(${\hat n}_1(\vec u; s)$, ${\hat n}_2(\vec u; s)$, ${\hat n}_3(\vec u; s)$)
attached to each point in phase space and requires that 
${\vec \xi} \cdot {\hat n}_3$ is a constant of motion. 
Clearly, ${\hat n}_3(\vec u; s)$ must be a solution of the T--BMT equation at 
$(\vec u; s)$
\footnote{The angle between two T--BMT solutions following  the same point 
on an orbit does not change in time. See also footnote 
{\it {\ref{foot:mont}}}.}. 
 At the same time the dreibein is chosen so that 
${\hat n}_i(\vec u; s)={\hat n}_i(\vec u; s + C)$ ( $i=1 \rightarrow 3$)
and so that
$\vec \xi$ precesses around ${\hat n}_3$ at a constant rate 
relative to  ${\hat n}_1$  and ${\hat n}_2$. The rate,
denoted by ${\check{\nu}}(\vec {\bar I})$,
should depend only on the actions  $\vec {\bar I}$.
The vectors  ${\hat n}_1$  and ${\hat n}_2$ are not solutions of the
T--BMT equation. 
The vector ${\hat n}_3$ has just the properties of the vector $\hat n$
introduced  earlier. 
This choice of the dreibein, which amounts to a $\vec u$ and $s$--dependent
rotation of the axes for describing the 
spin motion, is achieved by a suitably designed  
$\vec u$ and $s$--dependent canonical transformation which delivers a final
Hamiltonian (correct to first order in $\hbar$) with the 
`diagonalised' form \cite{yok86}
\begin{eqnarray}
{\check{h}} = \Sigma_i~ 2\pi {{\check Q}_i}{\check I}_i
          + 2\pi {\check{\nu}}{\check I}_{spin}
\end{eqnarray}
where 
${\check I}_{spin} = {\vec \xi} \cdot {\hat n}_3$ is
 now an integral of motion,
the spin action, and the ${{\check Q}}_i$ and ${\check I}_i$ 
are the corresponding orbital tunes and actions. 
The ${\check I}_i$ differ from the ${\bar I}_i$ by SG terms \cite{yok86}.
Note that the concept of spin tune has now been generalised; instead 
of the closed orbit spin tune ${\nu}_{spin}$
we have a spin tune 
${\check{\nu}}(\vec {\bar I})$ depending on the orbital actions (but not on 
${\check I}_{spin}$)
which differs slightly from  ${\nu}_{spin}$ and which 
reduces to ${\nu}_{spin}$ for zero orbital actions
\footnote{Note that for $\vec {\bar I} \neq {\vec 0}$,   
${\check{\nu}}(\vec {\bar I})$ {\it cannot}
normally be obtained from a complex eigenvalue of the naive one--turn 
eigenproblem discussed in
footnote {\it {\ref{foot:mont}}}.}.
Now, in retrospect, the
definition of resonance must be refined; we should really use
${\check{\nu}}(\vec {\bar I})$ in Eq.~(21) instead of ${\nu}_{spin}$.
It should now be clear why we sought a definition of spin precession rate,
i.e. spin tune, which makes the
latter independent of orbital phases and the azimuth $s$. Spin tune should
tell us something about the degree of long term coherence between the spin 
motion and the orbital motion and allow us to express this coherence by means
of resonance relations like Eq.~(21) (with ${\check{\nu}}(\vec {\bar I})$).
But if we work with a `fake spin tune' such as that obtained from the 
one--turn eigenproblem 
(see footnote {\it {\ref{foot:mont}}}~ and~ \cite{epac98})
 which 
depends on orbital phases so that the `fake spin phase advance' per turn 
varies from turn to turn, we can make no statements about long term coherence. 
With this redefinition of spin tune the dreibein
(${\hat n}_1$, ${\hat n}_2$, ${\hat n}_3$) is unique except at spin--orbit
resonances \cite{yok86,hh96}
and by this uniqueness the vector ${\hat n}_3$ is just the  vector
$\hat n$ introduced earlier except for a possible difference of sign!
The exotic (unstable) behaviour  of $\hat n$ near   
resonance is a manifestation of lack of uniqueness at resonance.

Now I return to the questions of smoothness and existence of a ${\hat n}$
obeying Eq. (18).
Since ${\check{\nu}}$ depends on orbital actions,  ${\hat n}$ is 
{\it potentially} nonunique 
at  almost all points in phase space because the resonance condition is 
satisfied almost everywhere if we include resonances of arbitrarily
high order. Thus ${\hat n}$ might not be differentiable in all directions in
phase space \cite{hei98}.
However, algorithms, both perturbative and nonperturbative, for 
constructing approximations to ${\hat n}$ are available \cite{br98}
(see also Article III ) and experience with calculating
${\hat n}$ by the author and colleagues 
seems to indicate that resonance effects rapidly become  weak as the resonance
order increases so that only a limited number of 
relatively low  order resonances are
likely to cause trouble. 
Therefore in the remainder of the article it will
be assumed that the spin field ${\hat n}$ is a legitimate tool in practice.
Nevertheless, the technical and interesting 
matters of existence and smoothness are under active
investigation \cite{khje98} and knowledge gained
from this study will be incorporated in our treatment of spin distributions.
An example of the dependence of ${\check{\nu}}$ on
amplitude can be found in \cite{epac98}. 

An extension of the numerical work 
reported in \cite{epac98} but carried out just before this article was 
completed indicates that ${\check{\nu}}$ actually 
`jumps over' resonant values as the orbit amplitude is changed 
\cite{vogt98,spin98}. That work is based on the `stroboscopic averaging'
algorithm in the computer code SPRINT \cite{hh96}. But even more recent 
results from a new version of the SODOM algorithm \cite{yok92,yok99}
corroborates these findings.
This implies, contrary to traditional 
expectations based on perturbation theory, that the spin--orbit resonance 
condition of Eq.~(21) is never exactly satisfied in non--perturbative
calculations. However, near to resonance, $\hat n$ still exhibits exotic
behaviour.

Although $\hat n$ and $\vec \xi$ both obey the T--BMT equation they are very  
different objects; $\hat n$ is a function of the dynamical phase space 
variables but $\vec \xi$ is a dynamical spin variable and by Eq.~(14) 
the Poisson bracket  $\lbrace {\hat n}, {\vec \xi}\rbrace$ vanishes.
Now that we have  a classical integral of motion for the spin, namely 
${\check I}_{spin}$, we  recognize  $\hat n$ as a phase space dependent 
 semiclassical quantisation axis corresponding to the quantum observable  
$(\hbar/2) \Psb \cdot \hat n$.
We also see that the quantisation axis coincides with the direction of the
equilibrium spin field. As we will see later $(\hbar/2) \Psb \cdot \hat n$
is a key  object in the analytical theory of
equilibrium electron polarisation and indeed it was originally introduced 
as an aid to calculating the electron polarisation \cite{dk72,dk73}.
The analysis becomes more complicated if the orbital motion is nonlinear
but in practice one tries to use an optic for which the nonlinear effects
have been minimised and tries to restrict the beam to a phase space volume
such that the motion is almost integrable.

One last point on the virtues of $\hat n$:  a calculation of electron 
polarisation
with the computer program SODOM \cite{yok92} which exploits  $\hat n$ 
agrees well with 
a calculation using the Monte--Carlo spin tracking program SITROS \cite{mb94}
which contains no notion of $\hat n$.

The material on spin distributions presented in this section is applicable 
both to electrons and protons. The application to protons is the topic of
Article III so that  for the remainder of this article  I will 
focus on electrons and in particular on the modifications
by synchrotron radiation to the concepts already presented.

\section{The effects of synchrotron radiation}
Synchrotron radiation  (SR) emitted by stored electrons has three main
effects: it determines the phase space distribution and it brings about
spin polarisation owing to  spin flip  associated with
synchrotron radiation (the ST effect) but the stochastic element of SR also
causes depolarisation.  Thus SR brings polarisation but it also takes it away!
As we have seen already and as we will  see below
spin motion is irrevocably intertwined with the orbital effects.
I will begin by summarising the orbital dynamics and  then discuss the
polarisation and depolarisation effects in detail.
 
\subsection{Orbital phase space}
Although SR spectra can be estimated by classical means \cite{jackbook}
SR is fundamentally a quantum phenomenon; it consists of
single photons so that one can only make reliable predictions by using 
quantised radiation theory. One then finds corrections to the classical
spectrum \cite{jack76}. The work of Huang and Ruth \cite{huang98} 
presented at this meeting is a good example of recent quantum calculations.

Most of the SR in conventional storage rings is generated in the fields of 
the dipole magnets defining the design orbit. 
A quantum  treatment for this case of the effects of SR on the orbital phase
space distribution  was carried out in 1975 \cite{dk75} using the Hamiltonian
of Eq.~(10). I will return to this later but here I will follow another
route which has the advantage of exhibiting the  transparency needed for this 
article.  

Photon emission in the dipole 
fields is largely incoherent and detailed calculations show that one can 
consider the photons to be emitted over short distances of the order of
$\rho/\gamma$  where $\rho$ is the orbit radius
\footnote{At this point I recommend
the reader to consult the chart of time scales for electron dynamics 
in \cite{mont84}. We
will need this on several occasions. Indeed, an appreciation of these time
scales is indispensible for understanding the physics of electron
storage rings.}.
Furthermore in practical storage rings the energy loss per turn of a single 
particle is small compared to the nominal energy. Thus the dissipative 
effect is weak and for example in HERA (Article II) an electron at
$27.5~GeV$ loses about $80~MeV$ per turn. 
Then for many purposes it  
suffices to describe the radiation reaction power ${\frak{p}}(s)$
from SR using a classical model in which smooth classical
radiation reaction power  ${\frak{p}}_{cl}(s)$ is overlayed with a 
`delta correlated'  (`white') stochastic
 component $\delta {\frak{p}}(s)$:
\begin{equation}
{\frak{p}}(s) = {\frak{p}}_{cl}(s) + \delta {\frak{p}}(s) \; , \;
\langle\delta {\frak{p}}(s)\delta {\frak{p}}(s')\rangle\;  =
  R(E_0,K)\delta (s-s')
\end{equation}
where the parameter $R$ quantifies the intensity of the noise and depends
on the design energy $E_0$ and the curvature $K$ \cite{bhmr91}.

The equations for $\vec u$ of 
deterministic orbital motion derived from a Hamiltonian are then modified by  
inclusion of damping and stochastic terms and  
in the (very good) approximation that the photons are emitted parallel
to the particle trajectory and neglecting interparticle interactions 
the resultant linearised
{\it stochastic differential equation} describing motion with respect 
to the closed orbit can be used to construct the Fokker--Planck
equation for the evolution of the phase space density 
\cite{bhmr91,jow87,rug90,risken}
\footnote{Restriction to linearised motion enables me to describe the chief 
qualitative features to be discerned without undue complication.}.
I write this as 
\begin{eqnarray}
\frac{\partial w} {\partial s}   =
       {\cal L}_{{}_{FP,orb}} \; w \; ,
\end{eqnarray}
where the orbital Fokker--Planck operator can be decomposed into the form:
\begin{eqnarray}
{\cal L}_{{}_{FP,orb}} = {\cal L}_{ham} + {\cal L}_{0} + {\cal L}_{1} + 
{\cal L}_{2} \, .
\nonumber
\end{eqnarray}

The term ${\cal L}_{ham} w$ is associated with the original symplectic 
(i.e. phase space density preserving) motion and it
contains just first order derivatives with respect to the components $u_i$
($i = 1...6$).
The operators ${\cal L}_{0}$ and ${\cal L}_{1}$ contain zeroth and first 
order derivatives and account for  damping  effects.
The operator ${\cal L}_{2}$
contains second order derivatives originating in diffusion effects.

A central  property of Eq.~(24) is that $w({\vec u};s)$
reaches equilibrium with 
$w(\vec u;s)= w(\vec u;s+C)$ within a few damping times.
At HERA at $27.5~GeV$ the longitudinal damping time is about
$7$ milliseconds 
$\approx 350$ turns $\approx$ (design energy)$/$(energy loss per turn) 
\cite{sands}.
Furthermore $w({\vec u};s)$ is 
gaussian and since the radiation effects 
are weak, $w({\vec u};s)$ is very close to being a solution of the
radiationless transport equation Eq.~(17) but with the radiation effects
determining the beam size and causing a tiny
ripple in the emittances
\footnote{For electrons I define the emittance of a mode to be the r.m.s.
action of the mode.} as functions of $s$.

Now that we understand the effects of SR on orbital phase space we can move 
on to spin.

\subsection{The Sokolov-Ternov effect}
Only a very small fraction of the radiated photons cause spin
flip but for electron spins aligned along a uniform magnetic
field, the $ \uparrow\downarrow $ and $\downarrow\uparrow$ flip rates
differ and this leads to a build-up of spin polarisation
antiparallel to the field. Positrons become polarised parallel to the field.
The transition rates for electrons are \cite{st64}:
\begin{eqnarray}
      W_{\uparrow\downarrow}
       &=&
      \frac{5\sqrt{3}}{16}
      \frac{e^{2}\gamma^{5}\hbar}
           {m_e^{2}c^{2}|\rho|^{3}}
      \left(1+
      \frac{8}{5\sqrt{3}}
          \right)
 \nonumber \\
      W_{\downarrow\uparrow}
       &=&
      \frac{5\sqrt{3}}{16}
      \frac{e^{2}\gamma^{5}\hbar}
           {m_e^{2}c^{2}|\rho|^{3}}
      \left(1-
      \frac{8}{5\sqrt{3}}
          \right)  \, .
\label{SokolovTernov}
\end{eqnarray}
For positrons, interchange plus and minus signs here and elsewhere.

The equilibrium polarisation in a uniform magnetic
field is independent of $\gamma$,
\begin{equation}
      P_{st} =
      \frac{
            W_{\uparrow\downarrow}
                    -
            W_{\downarrow\uparrow}
                                    }
           {
            W_{\uparrow\downarrow}
                    +
            W_{\downarrow\uparrow}
                                    }
      =
      \frac{8}{5\sqrt{3}}
     =
     92.38\%  \, .
\end{equation}
For a beam with zero initial polarisation,
the time dependence for build-up to equilibrium is
\begin{eqnarray}
    P(t)
       &=&
    P_{st}
    \left[
          1-\exp{(-t/\tau_{0})}
              \right]
\end{eqnarray}
where  the build-up rate is
\begin{equation}
    \tau_{0}^{-1}
      =
      \frac{5\sqrt{3}}{8}
      \frac{e^{2}\gamma^{5}\hbar}
           {m_e^{2}c^{2}|\rho|^{3}}  \, .
\label{eq:ST}
\end{equation}
The time $\tau_0$ depends strongly on $\gamma$ and $\rho$
but is typically minutes or hours.

However, the fields in storage rings are far from uniform 
but since the system is semiclassical, 
Eq.~(25), which was originally obtained from solutions of the 
Dirac equation, can be generalised 
and according to  Baier and Katkov\cite{bk68}
for electron spins initially aligned along an arbitrary
                                      unit vector $\hat\xi$ 
the transition rate is
\begin{equation}
      W
       =
      \frac{1}{2  \tau_{0}}
      \left[1-
      \frac{2}{9}(\hat{\xi}\cdot\hat{s})^{2}+
      \frac{8}{5\sqrt{3}}\,\hat{\xi}\cdot\hat{b}
          \right]
\end{equation}
where $\hat s$ = direction of motion and
$\hat b = ({\hat s} \wedge {\dot{\hat s}})/|{\dot{\hat s}}|$.
This is the magnetic field direction if the electric field vanishes
and the motion is perpendicular to the magnetic field.

The corresponding instantaneous rate of build-up of
                  polarisation along $\hat \xi$ is
\begin{equation}
      \tau^{-1}_{bk}
       =
      {\tau_{0}}^{-1}
      \left[1-
      \frac{2}{9}(\hat{\xi}\cdot\hat{s})^{2}
          \right] \, .
\label{eq:tauBK}
\end{equation}

But instead of spin flip rates we really need an EOM for the polarisation
itself and if we neglect 
the effect of stochastic (synchrotron radiation) photon emission
on the orbit and imagine that all particles remain on the closed
orbit (CO), the equation of motion for electron  polarisation 
as given by Baier, Katkov and Strakhovenko (BKS) is \cite{bks70,stck69}
\begin{eqnarray}
      \frac{d\vec{P}}{dt}
       &=&
      {\vec{\Omega}}_{co}\wedge\vec{P}
        -
      \frac{1}{\tau_{0}(s)}
      \left[
         \vec{P}-\frac{2}{9}
         \hat{s}
         (
           \vec{P}\cdot\hat{s}
                           )
               +
      \frac{8}{5\sqrt{3}}
      {\hat{b}}(s) 
          \right] \,   .
\end{eqnarray}
Note that the T--BMT term ${\vec{\Omega}}_{co}\wedge\vec{P}$ appears here
as the output of the radiation calculation itself.

By  noting that the characteristic time for polarisation build up is much 
larger than the circulation time
\footnote{Again, see \cite{mont84} for a compilation of time scales.},
and  integrating the BKS equation (Eq.~(31)) one finds the
generalised Sokolov--Ternov formula for the asymptotic electron polarisation
in arbitrary magnetic fields along the closed orbit \cite{mont84}:

\begin{equation}
         \vec{P}_{bks}
       =
       -  \frac{8}{5\,\sqrt{3}}~ \hat{n}_0 ~
     \frac{
   {\oint ds~
                {
                       ( {\hat{n}_{0}}(s)
                              \cdot
                        {\hat{b}}(s) )
                                } /
                     {|\rho(s)|^{3}}
                                    }
                                      }
          {
   {\oint ds~
                {
                            \left[
                                 1-
                             \frac{2}{9}
                       ({\hat{n}_{0}}(s)\cdot
                        \hat{s})^{2}
                                       \right]
                                                 } /
                    {|\rho(s)|^{3}}
                                    }
                                     } \, .
\label{GeneralisedST}
\end{equation}

So the polarisation settles down {\it aligned with ${\hat n}_0(s)$}, the 
periodic solution to the T-BMT equation on the closed orbit. 
In rings containing dipole spin rotators (Article II) the
polarization can usually not reach 92.38\% since ${\hat n}_0(s)$ is then not 
parallel to the field everywhere.
The corresponding polarisation build-up rate is
\begin{equation}
      \tau^{-1}_{bks}
       =
      \frac{5\sqrt{3}}{8}
      \frac{e^{2}\gamma^{5}\hbar}
           {m_e^{2}c^{2}}
      \frac{1}{C}\,
      \oint ds\,
                \frac{
                             \left[
                                 1-
                                 \frac{2}{9}\,
                       ({\hat{n}_0}\cdot
                        \hat{s})^{2}
                                       \right]
                                                 }
                     {|\rho(s)|^{3}}  \, .
\label{eq:tauBKS}
\end{equation}
\vspace{3mm}

\vspace{1mm}
The above formulae show that in the absence of stochastic motion 
the maximum attainable polarisation is 
$92.38\%$ instead of $100 \%$. Why should this be? 

At the simplest level the reason is clear: the probability for
reverse spin flip is nonzero (Eq.~(25)). Nevertheless, `lay observers' 
often imagine that spin flip has something to do with spin's trying to
reach the lowest energy level of the  two levels of a magnetic dipole in a 
magnetic field and that once the spin is in its lowest level it will stay 
there.
Then  $100 \%$ polarisation would be achieved. Also, reverse flip
 by radiation emission would defy energy conservation.

However, we are not dealing with spins at rest but with spins `sitting' on 
relativistic electrons which already have quantised orbital energy levels 
so that the prohibition of reverse flip by energy conservation no longer
applies.
{}From Eq.~(5) applied to electrons in a uniform vertical magnetic field it is 
clear that the energy change associated with spin reversal from up to down 
is $(1 + a \gamma =\gamma({1}/{\gamma} -1 + {g}/{2}))$
 larger than the separation of orbital energy levels 
$\hbar \omega_c$ where $\omega_c$ is the angular frequency of the
orbit. So one could naively imagine spin flip occuring without radiation but
simply by a change of orbital energy level. A related 
phenomenon involving exchange of orbital and spin energy
has been proposed by Derbenev \cite{derb90, bhr1} while commenting on the 
possible use of  transverse SG forces in storage rings. See Article III. 

Note also that the splitting of spin energy levels is not simply
proportional to ${g}/{2}$ but contains a Thomas precession term, which
indicates that the spin motion is coupled to the orbital motion.
Furthermore, the average energy of a synchrotron radiation photon is tens of
KeV. This is many orders of magnitude greater than the separation of
spin levels. Moreover, photons emitted during spin flip tend to have
higher energies than those emitted without spin flip.
 In addition, the  polarisation 
does not reverse its sign with respect to the magnetic field at 
$g = 0$ but at $g \approx 1.2$ \cite{dk73,jack76,bk68}.
This  results from  the
fact that $({1}/{\gamma} -1 + {g}/{2})$ appears in  the 
Hamiltonian $h_{int}^{dk}$ (Eq.~(10))instead of just ${g}/{2}$.
 
Finally, it is interesting to note that the synchrotron radiation spectrum
and the polarisation effects just depend on the curvature (i.e. the geometry) 
of the orbit \cite{bk68}. So the same effects could be obtained by using 
electric fields to bend the trajectory instead of magnetic fields.

These comments should convince the reader that 
in the laboratory frame we are not dealing with a 
simple two level spin system.  For further discussions relevant to this
topic the reader is directed to the articles by W. G. Unruh and J.D. Jackson
in these proceedings and elsewhere \cite{jack76}.

\subsection{Radiative depolarisation}

The stochastic element of photon emission
together with damping determines the equilibrium phase space
density distribution. The same photon emission also imparts a
stochastic element to the $\vec u$ in
$\vec{\tilde{\Omega}}({\vec u}; s)$  and then, via the T-BMT
equation applied to spin motion in the (inhomogeneous) focusing fields 
and in a simple classical picture, spin diffusion (and thus
depolarisation) can occur \cite{baiorl66}.
The  polarisation level reached 
is the result of a balance between the Sokolov--Ternov
effect and this radiative depolarisation. In practice, depolarisation can be
strong and it is therefore essential that it is well understood.

But how can we calculate the equilibrium polarisation?  After all, the 
polarisation at a point in phase space is the average of
the  unit spins $2{\vec \xi}/{\hbar}$ contained in a small packet of phase 
space at that point. Now, for protons, the phase space density is 
conserved along a trajectory so that no particles are lost from such 
a packet  but for electrons their stochastic motion means that 
spins are continually diffusing from packet to packet. For the orbital motion
one then employs a Fokker--Planck equation for the particle density. But 
polarisation is not a density so that it is not immediately clear how to 
proceed. Moreover the ST effect must be included so that an analogue of the
BKS expression for stochastic orbits is needed. 
I will mention the best  solution to this puzzle later but in the meantime 
I will follow a path which roughly reflects the way that estimates have
been made in practice.

A clue to the next step is contained in the above comments about the
equilibrium phase space distribution resulting from weak dissipation.
There, the phase space distribution settles down to a distribution close
to an invariant solution for the dissipationless problem but with the
widths of the distribution determined by the radiation.
Assuming that one has significant asymptotic
polarisation the characteristic depolarisation
time must be similar to the polarisation time, namely minutes or hours. 
Both are orders of magnitude larger than the orbital damping times. Thus 
the analogue for the polarisation would be that the direction of the 
equilibrium polarisation at each point in phase space would settle 
down close to the equilibrium solution of the radiationless problem,
namely ${\hat n}({\vec u};s)$. Furthermore, the `spin emittance' i.e. the 
average of ${\check I}_{spin} = {\vec \xi} \cdot {\hat n}$ at each point
in phase space, would be  independent of $\vec u $ and $s$.
 
As has been customary I will now adopt these plausible notions as 
working {\it assumptions} that at equilibrium a) the polarisation is parallel
to ${\hat n}({\vec u};s)$ and b) the {\it value} of the polarisation 
is independent of $\vec u $ and $s$. In particular, it is assumed that 
the spin tune hardly varies across phase space so that there are no `local'
spin--orbit resonances and therefore no polarisation `absorbers'.
 I will offer support for the first
assumption at the end of this article but in the meantime
some support for these assumptions comes from noting that by integrating the 
BKS equation along a deterministic synchrobetatron orbit
the polarization settles down very nearly parallel to  ${\hat n}$ \cite{us91}
in analogy with the solution on the closed orbit  (Eq.~{\ref{GeneralisedST}})
\footnote{`very nearly' means that the angle between the polarization and
${\hat n}$ is much less than the angle between $\hat n_0$ and
${\hat n}$.}.
Furthermore, a study of a special but  exactly solvable model of spin 
diffusion  \cite{kh97} shows that far from resonance the polarization
settles down asymptotically very nearly parallel to ${\hat n}$.

This picture was first proposed by Derbenev and Kondratenko \cite{dk73}.
In the absence of radiation $s_n = (\hbar/2) \Psb \cdot \hat n$ is conserved.
But in the presence of radiation one has 
\begin{eqnarray}
\frac{d s_n}{dt} = \frac{i}{\hbar}[h_{rad}^{dk} + h_{int}^{dk}, s_n] \, .
\end{eqnarray}

This is evaluated in the equations following Eq.~(4.2) in \cite{dk73}
and by writing $\vec {\frak s} =(\hbar/2) \Psb$
the essence of the physics can be stated (very) symbolically in the form: 
\begin{eqnarray}
\frac{d s_n}{dt} = 
\frac{d\vec{\frak s}}{dt}\cdot\vec n + 
                                  \vec {\frak s}\cdot\frac{d\vec n}{dt} \, .
\end{eqnarray}
The first term describes the rate of change of $s_n$ due to pure spin flip
at a point in phase space (pure ST effect). The consequent build--up of
polarisation is a `spin damping' analogous to orbital damping.
The second term describes the change in $s_n$
due to the fact that when a photon is emitted, the particle jumps 
without a change of spin to a new
position in phase space where it finds a new $\hat n$ which will in general
not be parallel to the $\hat n$ at the initial point. The projection of the 
spin on the $\hat n$--axis has thus decreased stochastically 
so that  $s_n$ diffuses in analogy with the diffusion of the orbital
actions. This is 
where the depolarisation comes in. Thus the effect on the 
polarisation of the stochastic journey of a particle though phase space
is accounted for by defining an appropriate quantisation axis at 
each point in phase space. Photon emission imparts both transverse and
longitudinal recoils to the electron but since a photon is emitted typically
within an angle $1/\gamma$ with respect to the direction of the electron, 
the effect of the longitudinal recoil (i.e. the energy jump) dominates:
the electron remains at almost the same point in $x$ and $z$ 
but can suffer a significant change in energy.
Then by neglecting the effect of transverse recoil 
Derbenev and Kondratenko arrive at the following expression for the 
equilibrium polarisation along the axes $\hat n$:
\begin{equation}
     {P}_{dk} =
        -\frac{8}{5\sqrt{3}}~
     \frac{
   { \oint ds \left< { \hat{b} \cdot ( \hat{n}- \frac{\partial{\hat{n}}}
                      {\partial{\delta}}
                           )}/{|\rho(s)|^{3}}
          \right>_{s}
                                    }
                                      }
          {
 {\oint ds  \left< ( { 1 -  \frac{2}{9} { ( \hat{n}\cdot\hat{s} )}^{2}
              + \frac{11}{18}
      \left( \frac{\partial{\hat{n}}} {\partial{\delta}}  \right)^{2}})/
                                                            { |\rho(s)|^{3}}
   \right>_{s}
                                    }
                                     }
\label{eq:PDK}
\end{equation}
where $\langle \ \rangle_{s}$ denotes an average over phase space
at azimuth $s$ and $\delta = \Delta E/E_0$ is the fractional energy deviation
from the design energy
\footnote{This is sometimes written as $\delta \gamma/\gamma$ \cite{dk73}.
See also footnote {\it {\ref{foot:gdndg}}}.}.
This formula differs from Eq.~(\ref{GeneralisedST}) by the
inclusion of the terms with  $\partial{\hat{n}}/{\partial{\delta}}$
and use of $\hat{n}$ instead of $\hat{n}_{0}$.
The derivative $\partial{\hat{n}}/{\partial{\delta}}$ is a measure of the
change of  $\hat{n}$ caused by fractional energy jumps $\delta$ and its
presence corresponds to the fact that the main consequence of a photon
emission is a change in particle energy.
The phase space  average of the polarisation is
\begin{equation}
  { \vec  P}_{av,{dk}}(s)
              =
     P_{dk}~
     \langle \hat{n} \rangle_{s}
\end{equation}
and $ \langle \hat{n}\rangle_{s}$ is very nearly aligned along
 ${{\hat n}_0}(s)$ (see the angle estimate below). The {\it value}
of the phase space  average,
 ${P}_{av,{dk}}(s)$, is essentially independent of $s$.

The effect of transverse recoil can also be included but contributes
derivative terms (see Article IV, Eq.~(2)) analogous to
${\partial{\hat{n}}}/{\partial{\delta}}$ which are typically
a factor $\gamma$ smaller than
${\partial{\hat{n}}}/{\partial{\delta}}$ and can usually be
neglected \cite{bm88,hs87}. This point will be dealt with  again
in Article IV .  

In the presence of radiative depolarisation Eq.~(\ref{eq:tauBKS}) becomes
\begin{eqnarray}
      \tau^{-1}_{dk}
       &=&
      \frac{5\sqrt{3}}{8}
      \frac{e^{2}\gamma^{5}\hbar}
           {m_e^{2}c^{2}}
      \frac{1}{C}
    ~\oint ds
                             \left<
                \frac{
                                 1-
                                 \frac{2}{9}
                       (\hat{n}\cdot
                        \hat{s})^{2}
                             +
                        \frac{11}{18}
                 \left(
                 \frac{\partial{\hat{n}}}
                      {\partial{\delta}}\right)^{2}
                                                 }
                     {|\rho(s)|^{3}}
                                       \right>_s \, .
\end{eqnarray}

Away from the spin--orbit resonances of Eq.~(21) 
$\hat{n}(\vec{u}; s) \approx {\hat n}_0(s)$. But near
resonances $\hat{n}(\vec{u}; s)$ deviates from ${\hat n}_0(s)$
by typically several 
tens of milliradians at a few tens of $GeV$ and the
deviation increases with distance in phase space  from the closed orbit.
The {\it spin orbit coupling function}
 ${\partial{\hat{n}}}/{\partial{\delta}}$, whose square 
$({\partial{\hat{n}}}/{\partial{\delta}})^2$ in Eq.~(36) 
quantifies the depolarisation,
can then be large and the equilibrium polarisation can then be small.
For example if
$|{\partial{\hat{n}}}/{\partial{\delta}}|$ is $1$ 
the polarisation will not exceed about $57 \%$.

Note that even very close to resonances,
   $|\langle \hat{n}\rangle_{s}| \approx 1$:
the phase space average polarisation measured by a polarimeter
is mainly influenced by the value of  $P_{dk}$ in Eq.~(37).

The nice thing about this formulation is that a very complicated calculation
of the effects of radiation has been distilled into a formula involving
a few strange coefficients (emanating from the radiation theory) and a 
classical solution to the T--BMT equation, namely $\hat n$ 
whose behaviour  encapsulates all of the depolarisation effects. 

To get high polarisation, one must have 
$(\partial{\hat{n}}/ \partial{\delta} )^{2} \ll 1$
in dipole magnets. If $\hat{n}$ is independent of the position in phase
space, the derivative is zero: all points in phase space have the
same quantisation axis  and there is no depolarisation. But storage ring 
fields are inhomogeneous so that $\hat n$ varies across phase space.
Thus the vector 
$\frac{\partial{\hat{n}}} {\partial{\delta}}$ depends on the optic of the
machine. The optimisation of the optic required to make 
${\partial{\hat{n}}}/{\partial{\delta}}$ small  is called
{\it spin matching} \cite{br98}.
This will be mentioned again in  Article II. An example can be found in
\cite{bonn90}.

The  term linear in  ${\partial{\hat{n}}}/{\partial{\delta}}$
in Eq.~(\ref{eq:PDK}) is due to a correlation between the spin orientation 
and the radiation power \cite{mont84,belom84}.
Alternatively, it can be considered to result 
from the interference between the two terms in Eq.~(35). 
In rings where ${\hat n}_0$ is horizontal due,
say, to the presence of a solenoid Siberian Snake \cite{spin961}, 
{}~${\partial{\hat{n}}}/{\partial{\delta}}$ has a vertical component
in the dipole fields. This can lead to a build-up of polarisation
({\it `kinetic polarisation'}) even though
the pure Sokolov--Ternov effect vanishes. The rate is $\tau^{-1}_{dk}$. 

The expression for $\tau^{-1}_{dk}$ in Eq.~(38) can be found from
a purely classical calculation of spin diffusion by evaluating 
the effects of the second term in Eq.~(35) or by other means
\cite{dk72,bhrnotes,bhmr91,br98} and indeed this was the first use for  
$\hat n$~\cite{dk72}. Then we have a mixed calculation: the spin flip is
described by quantum mechanics and the depolarisation is described by
classical diffusion.
But obviously kinetic polarisation will not be found
by that route and the exotic resonance structure examined in Article IV
would be missed. So it is clear that a quantum mechanical approach is needed
to get the full picture. An observation of kinetic polarisation \cite{spin961}
would be a nice vindication of this viewpoint.

The Derbenev-Kondratenko formula (Eq.~(36)) has been rederived in a  
very elegant way by Mane \cite{mane87}. He introduces the concept of 
generalised spin flip whereby he calculates the transition rates due to
photon emission from `spin up along \\
${\hat n}(x, p_x, z, p_z, \Delta t, \Delta E; s)$' to `spin down along  
${\hat n}(x, p_x, z, p_z, \Delta t, \Delta E - {\Delta}_{ph}; s)$'
where ${\Delta}_{ph}$ is the energy of the emitted photon. One also needs
the corresponding `spin down' to `spin up' rates. Then by requiring that
the polarisation has the same value over all of phase space and imposing
the constraint that the total generalised up--to--down rate equals the 
total down--to--up rate, and solving for the polarisation one arrives
at Eq.~(36)!  In this formulation, the concept of depolarisation never arises!
Instead one just has a statistical spin equilibrium.
In a perturbative calculation of 
$\langle({{\partial{\hat{n}}}/{\partial{\delta}}})^2 \rangle$ \cite{mane87}
one finds a series 
expansion in powers of emittances, i.e. in powers of $\hbar$, 
in which  each term contains a product of  
resonance denominators which impart the resonant structure to $\hat n$.
As the order of $\hbar$ increases along the series so does the order of the
 resonances contained in the successive terms. 

Apart from the calculation of Bell and Leinaas (Article IV),
there are two further quantum calculations which should be mentioned,
namely  those of Hand and Skuja \cite{hs87,hs89}. They choose ${\hat n}_0$
as the spin quantisation axis. 
When a photon is emitted, the electron jumps to a new orbit.
By writing the phase space coordinates as functions of the radiation fields
and including damping  phenomenologically,  they calculate 
the rates of spin flip along ${\hat n}_0$ \cite{hs87}
and  obtain an expression for the
equilibrium polarisation which is equivalent to the 
Derbenev--Kondratenko--Mane (DKM) expression in the limit in which the 
derivatives of $\hat n$ are evaluated in the linear 
approximation as in the SLIM 
formalism \cite{br98,chao81,chao82}.  Thus they only find the first order 
spin--orbit resonances \cite{br98}. Indeed, since their (quantum)
representation of the orbit has a form similar to the classical 
representation in \cite{yok82} their final expression
contains terms which are  equivalent and identical in form to terms 
obtained in \cite{yok82} in a model of classical spin diffusion. 
The fact that they only find first order resonances can then
be traced to their use of  just a first order perturbative
calculation and the choice of  ${\hat n}_0$ as 
quantisation axis. To find 
the higher order resonances one should use ${\hat n}$. However, in a second
calculation \cite{hs89}, again using ${\hat n}_0$, they calculated to higher 
order and although the outcome is not very transparent, the terms beyond the
leading  order in $\hbar$  contain high order resonant behaviour.  
But in the end  the moral seems to be that it is more efficient to choose a
quantisation axis (i.e. an unperturbed eigenstate), namely $\hat n$,
  which  reflects the physics and
invest the numerical effort in working with this.
In this respect the DKM formalism provides a clean 
practical framework in which to calculate higher order effects.
Radiation fields can cause spin flip. Now we can see how,  
by treating the external fields experienced by the spin as functions of
the radiation fields, spin precession can be regarded as spin flip as 
suggested at the beginning. 

We have seen that there are several ways to approach the estimation of the
equilibrium polarisation. 
In practice analytical calculations are carried out using the DKM formalism
and this has been a  successful and essential tool for predicting the main
qualitative features of polarisation in electron storage rings.
 However, I now present 
an approach which is in many ways more satisfying.

\subsection{Phase space and polarisation evolution equations}
Earlier, while discussing the difficulty of finding a Fokker--Planck
treatment for the polarisation I promised further insights and they follow 
now.

We have seen that the orbital phase space density
 $w$ obeys the Fokker--Planck equation, Eq.~(24).
Then if the ST effect is ignored and it is recognised that spin
is a passenger subject to the T--BMT equation it may be demonstrated
using a classical picture \cite{kh97,dbkh98}
that  the spin diffusion is described by the `Bloch' equation:
\begin{eqnarray}
\frac{\partial\vec{\cal P}} {\partial s}   =
     {\cal  L}_{{}_{FP,orb}} \; \vec{\cal P} +
\vec{{\tilde \Omega}} \wedge \vec{\cal P} \; ,
\end{eqnarray}
where  $\vec{{\tilde \Omega}}$ was defined by Eq.~(16) and 
$\vec{\cal P}$ is the 
{\it `polarisation density'} which is defined as
$2/\hbar \times $(density in phase space of spin angular momentum). 
The Bloch equation for the polarisation density is linear and it is 
universal in the
sense that it does not contain  the phase space density \cite{kh97}.
It is also valid far from spin--orbit equilibrium.
So the trick is to work with the polarisation density instead of the local 
polarisation ${\vec P}(\vec u; s)$.
It is simple to show that 
${\vec {\cal P}}(\vec u; s) = {\vec P}(\vec u; s){w}(\vec u; s)$
and then by combining Eqs.~(17) and (19) it is already clear that in the
absence of radiation, $\vec{\cal P}$ obeys the radiationless limit of Eq.~(39),
namely Eq.~(40) below.
The evolution equation for the local polarisation
in the presence of radiation  is obtained 
by combining Eq.~(24) with Eq.~(39). One finds that it
has a more complicated form than Eq.~(39) owing to
the presence of the second derivatives in
the ${\cal L}_{2}$ in ${\cal  L}_{{}_{FP,orb}}$. 
It also contains the phase space density so that it is not universal.
But the local polarisation can always be obtained instead as 
${\vec {\cal P}}(\vec u; s)/{w (\vec u; s)}$.

The ST effect can be included by
adding in terms from the BKS equation multiplied by the phase space density,
together with some terms to represent the interference between 
ST effect and diffusion\cite{kh97} and in fact the full Bloch equation
for the polarisation density and the Fokker--Planck  equation for the phase 
space density can be 
obtained from quantum radiation theory \cite{dk75}. 

In the absence of radiation we obtain:
\begin{equation}
  \frac{\partial {\vec{\cal P}}}{\partial s}~ =~
      \lbrace {\tilde h}_{orb},  {\vec{\cal P}}  \rbrace
        +   \vec{\tilde\Omega}  \wedge   {\vec{\cal P}} \; .
\end{equation}
{}From here it is easy to see, in analogy with  the case of the phase space, 
that since the orbital damping, orbital diffusion and ST terms
are very small compared to the remaining symplectic  and T--BMT terms, the
equilibrium (i.e. periodic)
${\vec P}(\vec u; s)$ will indeed be almost parallel to 
${\hat n}(\vec u; s)$, at least away from resonances
\footnote{The commentary in \cite{buon89} on our calculations
described in Article IV \cite{bm88}, contains the opinion that it is unsafe to
employ $\hat n$ as the appropriate quantisation axis. The above discussion
 should be an adequate response.}.

The Bloch equation for the polarisation density is free from assumptions
of the kind we needed earlier and in principle it allows us to calculate
everything we need from scratch 
by looking at the beam as a whole instead of focusing
on individual particles to begin with.
 It is clearly the best starting point
for discussing radiative polarisation.
Furthermore, the spin diffusion part (Eq.~(39)) can be set up for any 
source of noise in the orbital motion --- we just need the appropriate
${\cal  L}_{{}_{FP,orb}}$. For example it could be applied to 
scattering of protons by gas molecules.

\section*{Conclusion}
Spin polarisation in high energy storage rings is an exciting and exotic
topic.
I hope that the reader now has a solid overview of the status of our
understanding and will pass on to Articles II,~III,~IV and V. An overview 
of the experimental aspects of electron polarization and plans for the future 
can be found in \cite{spin962,ba98}.

\section*{Acknowledgments}
I would like to thank V.N. Baier, A.W. Chao, V. Balandin, Ya.S. Derbenev,
E. Gianfelice--Wendt, N. Golubeva,
K. Heinemann, G.H. Hoffst\"atter, A. Kondratenko,
H. Mais, S.R. Mane, G. Ripken, Yu.M Shatunov, M. Vogt
and K. Yokoya for their contributions to my understanding over a period of
many years and I thank K. Heinemann for his many valuable comments 
on this article.
Finally I would like to thank Pisin Chen for having the brilliant
idea of organising QABP98 and M. Berglund and T. Sen for careful reading of
the manuscript.

\end{document}